\documentclass[prl,twocolumn,citeautoscript,superscriptaddress]{revtex4}
\usepackage[dvips]{graphicx}
\usepackage{epsfig}
\usepackage{dcolumn}

\usepackage{amsmath}

\newcolumntype{e}{lrrrrr}

\begin{document}
\title{Theory of spin-conserving excitation of the $N-V^-$ center in diamond}

\author{Adam Gali} 
\affiliation{Department of Atomic Physics, Budapest
  University of Technology and Economics, Budafoki \'ut 8., H-1111, 
  Budapest, Hungary}
\affiliation{Department of Physics and School of 
Engineering and Applied Sciences, Harvard University, Cambridge, MA 02138, USA}

\author{Erik Janz\'en} \affiliation{Department of
  Physics, Chemistry and Biology, Link\"oping
  University, S-581 83 Link\"oping, Sweden}

\author{P\'eter De\'ak} \affiliation{Bremen Center for Computational
  Materials Science, Universit\"at Bremen, Am Fallturm 1, 28359 Bremen,
  German}

\author{Georg Kresse} \affiliation{Institut f\"ur Materialphysik,
  Universit\"at Wien, Sensengasse 8/12, 1090 Wien, Austria}

\author{Efthimios Kaxiras} \affiliation{Department of Physics and
  School of Engineering and Applied Sciences, Harvard University,
  Cambridge, MA 02138, USA}

\begin{abstract}
The negatively charged nitrogen-vacancy defect ($N-V^-$ center) in
  diamond is an important atomic-scale structure that can be used as a
  qubit in quantum computing and as a marker in biomedical
  applications. Its usefulness relies on the ability to optically
  excite electrons between well-defined gap states, which requires
  clear and detailed understanding of the relevant states and
  excitation processes. Here we show that by using hybrid
  density-functional-theory calculations in a large supercell we can
  reproduce the zero-phonon line and the Stokes and anti-Stokes
  shifts, yielding a complete picture of the spin-conserving
  excitation of this defect.
\end{abstract}


\maketitle
\begin{figure}
%\begin{center}
%\includegraphics[keepaspectratio,totalheight=6cm,width=8.0cm]{NVgeometel.eps}
\includegraphics[width=9.5cm]{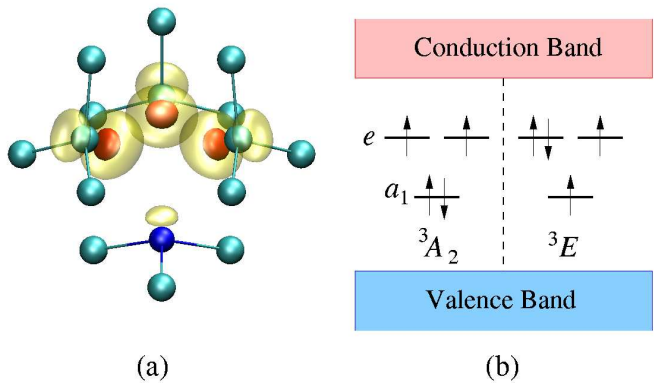}
%\end{center}
\caption{\label{fig:geometel}(Color online) (a) The structure of
$N-V^-$ center in diamond; only first- and second-neighbor C (cyan spheres) and
N (blue sphere) atoms to the vacant site are shown. 
The yellow and red lobes are contours of the calculated difference in
spin density for the $^3E$ state as obtained by
 the HSE06 and PBE functionals.
(b) Schematic diagram of the defect states in the gap 
and their occupation in the $^3A_2$
(ground) and $^3E$ (excited) states. }
\end{figure}

%\section{Introduction}
Quantum computing and its many exciting applications relies on the
successful realization of quantum logic bits (qubits) that can operate
under practically feasible conditions.  Few physical systems can meet
the requirements of controlled quantum coherence and robust
operational conditions.  One of the most promising candidates is the
negatively charged nitrogen-vacancy defect ($N-V^-$ center) in bulk
diamond \cite{duPreez65,Davies76}: the spin state of this defect can
be manipulated using excitation from the $^3A_2$ ground state to the
$^3E$ excited state by optical absorption (Fig.~\ref{fig:geometel}).
The main advantage of the $N-V^-$ center is that it can operate at
\emph{room temperature} as a solid state qubit
\cite{Wrachtrup01,Jelezko04-2,Epstein05,Hanson06PRL,Childress06,Childress07,Hanson08}.
Detailed understanding of this excitation process is crucial in the
realization of qubits based on diamond.  However, achieving this level
of understanding stretches the capabilities of theoretical methods
that are usually applied to the study of defects in solids because of
the special nature of the $N-V^-$ defect.  Specifically, this defect
combines strong coupling of ionic and electronic degrees of freedom
with a many-body character of the electronic states, and its
interpretation is further complicated by contradictory experimental
measurements of the emission spectrum at different temperatures
\cite{Davies76,Manson05}.

In this Letter, we report a theoretical investigation of the radiative
transitions of the $N-V^-$ center using electronic structure
calculations, which give a consistent and accurate account of the
excitations observed and provide a plausible resolution of the
experimental situation.  We utilized the HSE06 screened Hartree--Fock
hybrid exchange-correlation density functional
\cite{heyd:8207,krukau:224106} to determine the geometry and
excitation energies of the $N-V^-$ center, and we compare the results
to the traditional PBE \cite{PBE} exchange-correlation density
functional and the experimental data. We find that -- in contrast to
PBE -- the HSE06 functional (which reproduces the band gap of diamond
within 0.5\%), can also reproduce both the zero-phonon line and the
Stokes-shift quantitatively (within 1.5\% in this case).  This result
demonstrates that hybrid functionals improve not only the excitation
of the extended system (gap) but also localized ones. This promises a
very significant advantage in defect calculations. Motivated by the
success of the hybrid functional to reproduce key experimental
measurements of the excitation process, we calculate an anti-Stokes
shift of 0.217 eV, measured to be 0.185~eV at usual experimental
conditions at low temperature~\cite{Davies76}.  From this result, we
argue that the anti-Stokes shift of 0.065~eV measured at low
temperature under high-energy-density laser
illumination~\cite{Manson05}, is most likely due to the local heating
of the sample caused by the focused laser beam.

Previous density functional theory (DFT) calculations have
shown \cite{Goss96,Gali08,Larsson08,hossain:226403} that well defined
defect levels appear in the band gap due to this defect: a fully
occupied $a_1$ level and a doubly occupied two-fold degenerate
$e$-level at a higher energy. The electrons have parallel spins on
the $e$-level preserving the C$_{3v}$ symmetry of the defect; 
a many-body wavefunction built from single-particle states in 
the gap represents the $^3A_2$ ground state. 
The excitation can be understood as
promoting one electron from the $a_1$ level to the $e$ level resulting
in a new many-body excited state, $^3E$, as shown 
schematically in Fig.~\ref{fig:geometel}.

\begin{figure}
\includegraphics[width=2.5in]{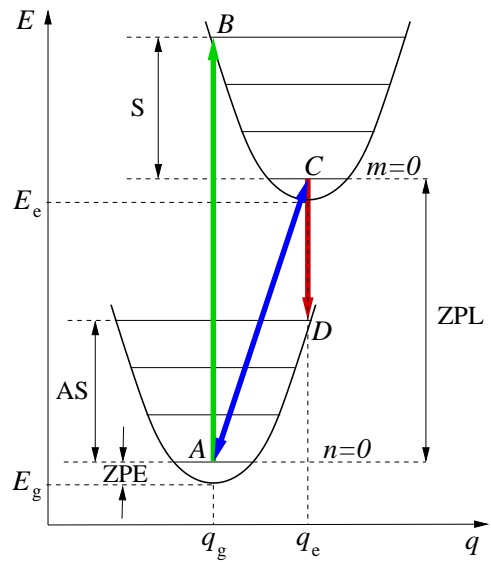}
\caption{\label{fig:confcoord}(Color online) The energy ($E$) vs. configuration
  coordinate ($q$) diagram for the excitation process of a defect
  in the Franck-Condon approximation:
  $E_\text{g}$, $E_\text{e}$  are the minima in the quasi-parabolic
  potential energy surfaces of the defect in the ground and excited
  states, respectively and 
  $q_\text{g}$, $q_\text{e}$ the corresponding coordinates. ZPE is the
  zero point energy (indicated only for the ground state). The energy
  ladders show the phonon energies with the phonon ground states at
  $n=0$ (ground state of the defect) and $m=0$ (excited state). At elevated
  temperatures the high-energy phonon states can be occupied by
  inducing transition $A \rightarrow B$ (vertical absorption, green arrow), and
  $C \rightarrow D$ (vertical emission, red arrow). Transition
  $A \leftrightarrow C$ corresponds to the zero-phonon line (ZPL, blue double arrow) both in
  absorption and emission. The energy of the Stokes shift (S) and
  anti-Stokes shift (AS) are also shown.
  }
\end{figure}

Excitation changes not only the electron wavefunction but the atomic
structure of the defect as well. Hence, the ground state and the
excited state will possess different potential energy surfaces (PES)
and different vibrational states, as shown schematically in
Fig.~\ref{fig:confcoord}. The transition between the lowest PES will
result in the zero-phonon line (ZPL) both in absorption and emission,
a process in which no real phonons are involved in the excitation or
de-excitation process. The ZPL was measured at 1.945~eV (yellow light)
both in low temperature absorption and emission~\cite{Davies76}. At
liquid-nitrogen temperature a broad phonon side band was measured in
absorption with phonon-related peaks at approximately 2.020, 2.110,
2.180 and 2.250~eV, with the highest intensity at 2.180~eV (green
light), while in the emission band the first phonon sidebands, better
resolved than in absorption, are found at 1.880, 1.820, and 1.760~eV,
with the 1.760~eV peak (red light) having the highest intensity
\cite{Davies76}.

The Franck--Condon approximation is commonly used to interpret the
excitation spectrum, that is, assuming that the electronic transition
is very fast compared with the motion of nuclei in the lattice.  In
addition to the Franck--Condon assumption, three other approximations
are commonly assumed (see Fig.~\ref{fig:confcoord}). The first is that
each lattice vibrational mode is well described by a quantum harmonic
oscillator, as implied by the quasi-parabolic shape of the potential
wells, and almost constant energy spacing between phonon energy
levels. The second, called the low-temperature approximation, is that
only the lowest (zero-point) lattice vibration is excited, implying
that electronic transitions do not originate from any of the higher
phonon levels. The third, called the linear-coupling approximation, is
that the interaction between the defect and the lattice is the same in
both the ground and the excited states; this implies two equally
shaped parabolic potentials and equally spaced phonon energy levels in
both the ground and excited states. The detected phonon peaks in the
spectrum may be associated with the $m=1,2,\dots$ and $n=1,2,\dots$
quanta of a characteristic phonon mode in absorption and emission.

The highest intensity in the phonon side band at low temperatures
corresponds to the excitations where the geometry does not change,
that is, the vertical absorption and emission~\cite{Huang50}.  This
way the Stokes and anti-Stokes shifts are determined, which reveal the
relaxation energy of the atoms due to the electronic excitation (see
Fig.~\ref{fig:confcoord}). In the linear coupling approximation the
Stokes and anti-Stokes shifts would have the same value but
experimental measurements indicate a difference of about 0.05~eV (see
Table~\ref{tab:exc} and related discussion).

Another complication in understanding the excitation and de-excitation
processes arises from a recent measurement~\cite{Manson05} of the
highest emission intensity in the phonon sideband at
$\approx$1.880~eV.  In this measurement, the second peak at 1.820~eV
is clearly visible but the third at 1.760~eV is almost
missing~\cite{Manson05}. This measurement was carried out at 10~K.
This is puzzling since at higher temperature (for instance, at liquid
nitrogen temperature, 77~K, where earlier experiments were
conducted~\cite{Davies76}), higher-energy phonon states can be
occupied resulting in larger vertical emission energy, the opposite of
what was found in the recent measurements~\cite{Manson05}.  We note
that in the latter, low-temperature (10~K) experiment, a laser beam
was focused with a 10~cm lens, increasing the energy density by a
factor of 10$^4$. In fact, this experiment measured the ionization of
$N-V^-$ defects from the $^3E$ excited state during the high intense
excitation~\cite{Manson05}.  Assuming that all the usual
approximations hold for this process, an anti-Stokes shift of 0.065~eV
can be deduced~\cite{Manson05} which clearly contradicts the results
of earlier experiments~\cite{Davies76}.  This issue must be addressed
and resolved in a full explanation of the process.

%\section{Computational methodology}

State-of-the-art methods to investigate defect properties in solids
employ DFT in a supercell geometry. While the calculation of the
ground state charge and spin density can be obtained accurately using
the local density approximation (LDA) for the exchange-correlation
functional or by other functionals that include density-gradient
corrections (for example, the PBE functional~\cite{PBE}), the accurate
calculation of the excitation energies presents a challenging problem
due to the well-known self-interaction error of these methods. For
example, the LDA value for the zero-phonon line (ZPL) of the $N-V^-$
center is 1.71~eV~\cite{Gali08} compared to the experimental data,
1.945~eV \cite{Davies76}. The Bethe-Salpeter equation followed by the
parameter-free GW-method for the quasi-particle correction of the
Kohn-Sham levels~\cite{hedin:1969} is the best tool to calculate the
excitation energies, but it is computationally prohibitive for large
supercells. Recently, it was shown that both the band
gap~\cite{marsman:064201} and the excitation energies
\cite{paier:121201} are vastly improved by applying a screened
Hartree--Fock hybrid density
functional~\cite{heyd:8207,krukau:224106}, referred to as the HSE06
hybrid functional.

We employed two methods to calculate the excitation properties of the
$N-V^-$ center in diamond: the traditional PBE functional and the
HSE06 hybrid functional. First, the diamond primitive lattice was
optimized, then a simple cubic 512-atom supercell was constructed.
Finally, we placed the negatively charged nitrogen-vacancy defect in
the supercell, and optimized the structure for each given electronic
configuration.  We employed the {\sc VASP5.1} code~\cite{Kresse96} to
carry out these calculations with a plane-wave basis set (using an
energy cut-off of 420~eV) and PAW-type potentials to model the atomic
cores~\cite{Blochl94,Kresse99}.  For the optimization of the lattice constant
we used a plane-wave cut-off energy of 840~eV and a $12 \times 12
\times 12$ Monkhorst--Pack k-point set~\cite{MP76} for the primitive
diamond lattice.  For the 512-atom supercell we used the
$\Gamma$-point that provides a well-converged charge density.  We note
here that all these calculations use the Born--Oppenheimer
approximation, so the electron states are calculated as a function of
the coordinates of the nuclei treated as classical particles. This
approximation is valid at low temperatures.

%\section{Results on the perfect diamond}

The calculated lattice constant of diamond is slightly different in
the PBE (3.567~\AA) and the HSE06 (3.545~\AA) approximations. The
calculated band gap is very different in the two approaches: the PBE
functional yields 4.16~eV, while the HSE06 functional gives 5.43~eV,
very close to the experimental value of 5.48~eV.  The PBE-functional
error is large both in absolute value (1.32~eV) and relative value
(24.1\%); these values are reduced by the HSE06 functional to 0.05~eV
and 1.0\%, respectively.  We expect that the calculated excitation
energies of the defect excited states will be also improved by using
hybrid functionals~\cite{Deak05,alkauskas:046405,Deak08}.

%\section{The electronic states of the NV center in diamond}

The $^3A_2$ ground state of the $N-V^-$ center is obtained by spin
polarized calculations both with the PBE and HSE06 functionals. The
$^3E$ excited state is simply obtained by promoting one electron from
the $a_1$ defect level to the $e$ defect level in the band
gap~\cite{Goss96,Gali08}. In the {\sc VASP} code it is possible to set
the occupation numbers of the single particle levels, thus the $^3E$
excited state can be calculated in a self-consistent manner through
such constrained occupation.  The total energy was minimized for both
electronic configurations as a function of the coordinates of the
nuclei, which allows us to determine the configuration coordinates of
the ground state ($q_\text{g}$) and the excited state ($q_\text{e}$)
of the defect.  The corresponding energy minima in the calculations
are shown as $E_\text{g}$ and $E_\text{e}$ in
Fig.~\ref{fig:confcoord}. The zero-point vibration states ($n=0$ and
$m=0$) will raise these energies by a value of order a few tens meV,
called zero-point energy (ZPE, shown in Fig.~\ref{fig:confcoord}); for
example, Davies and Hamer~\cite{Davies76} deduced a ZPE value of
$\approx$35 meV. We note that the difference between the ZPE values of
$E_\text{g}$ and $E_\text{e}$ is expected to be even smaller, of order
a few meV. The energy difference between the energy minima of
$E_\text{g}$ and $E_\text{e}$ therefore gives a very good estimate of
the ZPL ($A\rightarrow C$ transition, see Fig.
\ref{fig:confcoord}). The transitions $A\rightarrow B$ and
$C\rightarrow D$ are readily calculated by fixing the geometry at
$q_\text{g}$ and $q_\text{e}$, respectively, while varying the
electronic configurations as explained above. We note that the error
associated with the ZPE cannot be avoided in the calculated
$A\rightarrow B$ and $C\rightarrow D$ transitions, which means that
the values for these transitions are less accurate than for the ZPL.
The calculated excitation energies using the PBE and HSE06 functionals
are given in Table~\ref{tab:exc}.

\begin{table}
  \caption{\label{tab:exc} The calculated vertical absorption
    ($A\rightarrow B$) and vertical emission energies ($C\rightarrow D$),
   and the zero-phonon line (ZPL) obtained with the 
   PBE and HSE06 functionals,
    compared to measured values from
    Ref.~[\onlinecite{Davies76}]. The Stokes-shift (S) between the
    vertical absorption and the ZPL, and the anti-Stokes-shift (AS)
    between the ZPL and the vertical emission are also given (all values in eV).}
\begin{ruledtabular}
\begin{tabular}{e}
       &  \multicolumn{1}{c}{ZPL}   &  
          \multicolumn{1}{c}{$A\rightarrow B$} &  
          \multicolumn{1}{c}{S} & 
          \multicolumn{1}{c}{$C\rightarrow D$}   & 
          \multicolumn{1}{c}{AS} \\ 
\hline
PBE    &  1.706 &     1.910         &  0.204      &    1.534  &  0.172 \\
HSE06  &  1.955 &     2.213         &  0.258      &    1.738  &  0.217 \\ 
Exp.~[\onlinecite{Davies76}]   &  1.945 &     2.180         &  0.235      &    1.760  &  0.185 \\     
\end{tabular}
\end{ruledtabular}
\end{table}

%\section{Discussion}

The PBE functional gives too low value for the ZPL ($\approx$1.71~eV),
so this gradient corrected functional does not improve the LDA value
(1.71~eV \cite{Gali08}) at all. This is not surprising since both the
LDA and the PBE functionals suffer from the self-interaction
error. However, the HSE06 functional gives an almost perfect value,
the difference from experiment being smaller than 0.5\%. Apparently,
the HSE06 functional improves not just the band gap of the perfect
semiconductor but the defect internal transition energy as well.  The
PBE functional does not improve the LDA values for the vertical
absorption energy either, which is again too low by $\approx$0.3~eV.
The HSE06 functional yields an almost perfect value for the energy of
the vertical absorption compared to the experimental result (within
1.4\%). The larger discrepancy for the $A\rightarrow B$ transition
than for the ZPL may be attributed to the intrinsic ZPE error as
explained above.  We note that the calculated Stokes shift (the
relaxation energy), is close to the experimental result from both the
PBE and the HSE06 functionals. The reason is that the self-interaction
error inherent in the PBE functional is almost fully canceled, as the
relaxation energy corresponds to the energy difference between two
different atomic configurations with the same electronic
configuration.  The wavefunctions and the spin density obtained with
the HSE06 functional are somewhat more localized than those obtained
with the PBE functional. For example, the integrated spin density of
the $^3E$ state in a 5$^3$~\AA $^3$ cube centered at the vacant site
containing the three carbon atoms and the nitrogen atom is 1.61 and
1.64 obtained in PBE and HSE06 calculations, respectively (see
Fig.~\ref{fig:geometel}). This is expected of the HSE06 functional
which contains the Hartree--Fock exchange, giving more localized wave
functions than the pure DFT-PBE.

Having established the high level of accuracy of the HSE06 functional,
we can address the issue of the vertical emission energy and the
anti-Stokes shift. The value calculated with the PBE functional is
again very low and not comparable to any experimental data. However,
the HSE06 value (1.738~eV) is very close to the measurement of Davies
and Hamer (1.760~eV) \cite{Davies76}. Since our calculations are valid
at low temperature we assume that the low temperature approximation
beyond the Franck--Condon assumption holds for this transition.  We
conclude that the anti-Stokes shift is 0.185~eV at usual experimental
conditions at low temperatures for the $N-V^-$ defect in diamond. For
this defect the linear-coupling approximation does not hold. The
difference between the Stokes and anti-Stoke shifts is
$\approx$(0.235-0.185)~eV=0.050~eV in experiment, which compares
favorably to the HSE06 value (0.258-0.217)~eV=0.041~eV. This indicates
somewhat different shape of the PES for the ground state and the
excited state, thus different vibration modes.

The only remaining unresolved issue is the recent experiment
suggesting a much lower anti-Stokes value of 0.065 eV~\cite{Manson05}.
We suggest that the high energy density excitation in this experiment
resulted in local heating of the sample in the area where the laser
beam was focused. The local heating of the sample will break the low
temperature approximation and will cause a shift in the occupation of
the phonon states from $m=0$ to $m=2$. That may explain why the
detected vertical emission energy is larger at 10~K (at high energy
density) than at 77~K (at low energy density) excitation. Our results
indicate that detailed analysis of the vibration modes of the $^3E$
excited state is important for a complete understanding of the
radiative emission of the $N-V^-$ center; further experimental and
theoretical efforts are needed in this direction.

%\section{Summary and Conclusions}
%In summary, we showed that the internal transition energies of the
%$N-V^-$ center in diamond can be quantitatively determined by HSE06
%hybrid functional. Our calculations indicate that high energy density
%excitation leads to luminescence from the high vibration states even
%at very low temperatures. 

%\section*{Acknowledgments}
AG acknowledges support from Hungarian OTKA No.\ K-67886 and the grant
of SNIC001-08-175 from the Swedish National Supercomputer Center; we thank 
Jeronimo Maze for fruitful discussions. 

%\bibliographystyle{prsty4}
%\bibliography{references}

\end{document}